\documentclass[usegraphicx,usenatbib]{mn2e}
\begin{document}

\title{A Dark Galaxy in the Virgo Cluster Imaged at 21-cm.}

\author[R. F. Minchin et al.]{R. F. Minchin$^{1}$,
J. I. Davies$^{1}$,
M. J. Disney$^{1}$,
A. R. Marble$^{2}$,
C. D. Impey$^{2}$,
\newauthor
P. J. Boyce$^{3}$,
D. A. Garcia$^{1}$,
M. Grossi$^{1}$,
C. A. Jordan$^{4}$,
R. H. Lang$^{1}$,
\newauthor
S. Roberts$^{1}$,
S. Sabatini$^{5}$,
W. van Driel$^{6}$
\\
$^{1}$ School of Physics and Astronomy, Cardiff University, Cardiff,
CF24 3YB, UK\\
$^{2}$ Steward Observatory, University of Arizona, 933 N. Cherry Ave., Tucson,
AZ 85721-0065, US\\
$^{3}$ Planning Division, Cardiff University, Park Place, Cardiff,
CF10 3UA, UK\\
$^{4}$ Jodrell Bank Observatory, University of Manchester,
Macclesfield, Cheshire, SK11 9DL, UK\\
$^{5}$ Osservatorio Astronomico di Roma, via Frascati 33, I-00040,
Monte Porzio, Italy\\
$^{6}$ Observatoire de Paris, GEPI, CNRS UMR 8111 and Universit\'e
Paris 7, 5 place Jules Janssen, F-92195 Meudon Cedex, France}
\maketitle

\begin{abstract}
Dark Matter supposedly dominates the extragalactic, yet no totally dark
structure of galactic proportions has ever been convincingly identified.
Earlier \citep{Minchin05} we suggested that VIRGOHI 21, 
a 21-cm source we found in the Virgo Cluster at Jodrell Bank using single-dish
observations \citep{Davies04}, was probably such a dark galaxy because of its 
broad line-width ($\sim 200$ km\,s$^{-1}$) unaccompanied by any visible
gravitational source to account for it.  Now we have managed to image VIRGOHI
21 in the neutral-hydrogen line, and indeed we find what appears to be a dark,
edge-on, spinning disc with the mass and diameter of a typical spiral galaxy.
Moreover the disc has unquestionably interacted with NGC 4254, a luminous
spiral with an odd one-armed morphology, but lacking the massive interactor
invariably responsible for such a feature.  Published numerical 
models \citep*{Vollmer05} of NGC 4254 call for a close interaction $\sim 10^8$ 
years ago with a perturber of $\sim 10^{11}$ solar masses.  This we take as 
completely independent evidence for the massive nature of VIRGOHI 21.
\end{abstract}


\section{Introduction}

Cold Dark Matter (CDM) models of galaxy formation predict many more dark
matter halos than are observed as galaxies 
\citep*{Kauffmann93,Moore99,DOnghia04}.
Whether stars can form in dark matter halos depends critically on
the fraction of baryons ($m_{d}$) that can be trapped and retained by
each halo (for previous discussions about the existence and formation of
dark galaxies see \citep{Jimenez97,Hawkins97, Verde02}). Some of those
that retain a small fraction of their original baryons ($m_{d}<0.05$)
are able to form stable gaseus disks, but because of the low gas
densities they are not able to form stars. The physical state of the gas
in the disc is very dependent on its density and temperature with the
latter influenced by the intensity of any ionising background. Models
predict that galaxies can form with gas column densities that prohibit
star formation yet provide some self shielding from the ionizing
background (Davies et al. in prep.). Such dark galaxies are potentially
detectable by blind 21cm surveys of the sky.

Objects detected at 21cm but with no optical counterparts have been
known about for many years; these include high velocity clouds
\citep{Wakker97}, the Leo Ring \citep{Schneider83}, and various gas
clouds
close to bright galaxies \citep{Kilborn00,Boyce01,Ryder01}.  However,
none of these objects have the characteristics of a galaxy, i.e.
detectable emission over galaxy sized spatial scales and a velocity
structure consistent with a rotating and gravitationally bound disc.

In a previous paper \citep{Minchin05} we described Jodrell Bank and
Arecibo observations of VIRGOHI 21 an HI source discovered during a
survey of the Virgo cluster \citep{Davies04}. VIRGOHI 21 has all of the
HI characteristics of a typical rotating disc galaxy, but no detectable
optical emission. Our conclusion was that VIRGOHI 21 was an extremely
promising candidate for the first dark galaxy. In this paper, we present
high resolution HI observations of this
source, which add to the conclusions drawn in the earlier paper that
this is, indeed, a massive dark galaxy.

\section{Observations and Analysis}

The new data were taken in March 2005 at the Westerbork Synthesis Radio 
Telescope (WSRT) in two full 12-hour syntheses and reduced using the
{\sc Miriad} package.  The data were flagged for shadowing and on two of the 
fourteen 25-metre antennae one polarisation was flagged due 
to problems with the gain.  A spectral bandwidth of 10 MHz covered the velocity
range 930 - 3070 km\,s$^{-1}$.  Removal of the noisier end channels left 230 
useful channels of width 8.2 km\,s$^{-1}$ each, giving a velocity resolution 
of 10 km\,s$^{-1}$ over the range 980 to 2890 km\,s$^{-1}$.  Continuum removal 
was carried out in the UV plane using {\sc Uvlin}.  The standard source 3C147
was used for calibration.  Cleaning used a robust setting of 1,
close to normal weighting \citep{Briggs95}. The cleaned cube was gaussian 
smoothed spatially and Hanning smoothed in velocity, and then used for a 
second-pass deeper cleaning which gave the final cube used in the analysis.  
The synthesised beam was $99\arcsec\times 30\arcsec$ in size 
(extended North-South) and the noise was 0.3 mJy\,beam$^{-1}$\,channel$^{-1}$, 
giving a $5\sigma$ column-density to sources 25 km\,s$^{-1}$ wide of $2\times
10^{19}$ Hydrogen atoms cm$^{-2}$.

\begin{figure*}
\begin{minipage}{126mm}
\includegraphics[width=126mm]{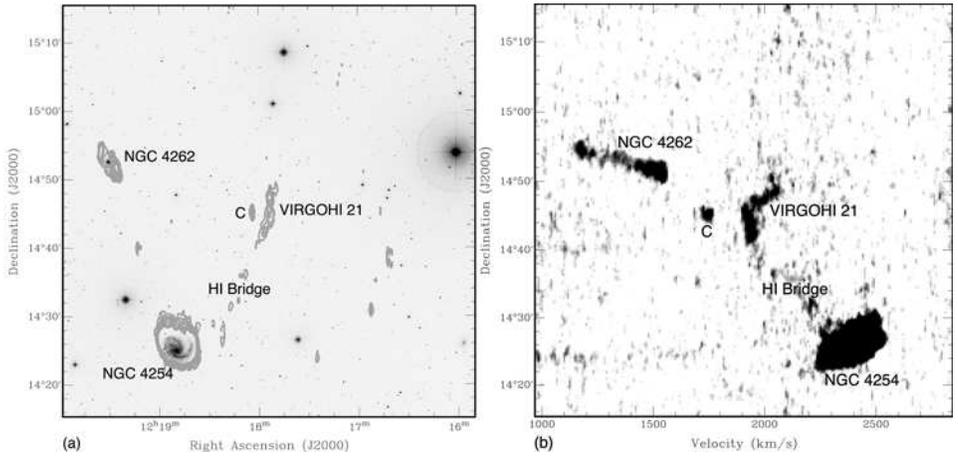}
\caption{(a) H\,{\sc i} contour map of the 21-cm observations, superimposed
on a 1 square degree negative Digitized Sky Survey image. Contours are from
$2.5 \times 10^{19}$ to $2 \times 10^{20}$ cm$^{-2}$ at intervals of $2.5 
\times 10^{19}$ cm$^{-2}$.(b) Shows the declination-velocity projection of 
the data cube.  More detail can be seen in the on-line animation of the
cube.}
\end{minipage}
\end{figure*}

Fig. 1a shows a neutral Hydrogen (H\,{\sc i}) contour map of the field 
superimposed on a negative optical image.  VIRGOHI 21 is the elongated 
structure in the centre (which is at about 2000 km\,s$^{-1}$).  A faint bridge 
can be seen stretching down to the prominent spiral NGC 4254 
(2400 km\,s$^{-1}$) while the other two sources, NGC 4262 (1500 km\,s$^{-1}$, 
upper left) and the faint galaxy `C' (1750 km\,s$^{-1}$, immediately to the 
left of VIRGOHI 21) appear unconnected.  Fig. 1b shows the velocity-declination
projection of the full 3 dimensional data cube.  Now NGC 4254 is at the bottom 
right while VIRGOHI 21 is the angular structure in the centre. C is to its 
left.  Far more detail can be seen on an animation which is available
at http://www.astro.cf.ac.uk/groups/galaxies, for instance the bridge is clear 
and obvious, as is the lack of any connection between VIRGOHI 21 and either
 NGC 4262 or C.  The apparent alignment of NGC 4262 with VIRGOHI 21 and NGC 
4254 in Fig. 1b is a consequence of the particular projection shown

The  $\sim$ 25 arcmin filamentary bridge stretches from the low-velocity 
(western) edge of NGC 4254, falling gently in radial velocity from 2250 
km\,s$^{-1}$ at declination $+14^\circ
20^\prime$ towards 1900 km\,s$^{-1}$ at $+14^\circ 41^\prime$ where it is 
suddenly arrested.  Then at $+14^\circ 46^\prime$ it is abruptly
wrenched upwards again towards 2100 km\,s$^{-1}$ at $14^\circ 49^\prime$.

Fig. 2 is a blow-up of the source region superimposed on a far deeper CCD
optical image, illustrating that there is no optical counterpart 
\citep{Minchin05}.  An optical spectrum from the 6.5-m MMT in Arizona of the 
small, faint galaxy `A', which is superposed on the highest H\,{\sc i}
contour ($1\times 10^{20}$ cm$^{-2}$) at declination $+14^\circ47.4^\prime$,
shows that it is at a redshift of $z=0.25$ and is therefore unconnected 
with VIRGOHI 21.  The 17th magnitude galaxy `C' to the left at 
$+14^\circ45^\prime$ is an H\,{\sc i} point-source at this resolution.  By 
comparison, VIRGOHI 21 is an extended structure in both dimensions rather than 
a collection of discrete compact clouds.  The velocity-declination plot (right)
shows the complex kinematic structure of VIRGOHI 21.  The most remarkable
feature is the tilted portion between $+14^\circ46^\prime$ (1900 km\,s$^{-1}$) 
and $+14^\circ49^\prime$ (2100 km\,s$^{-1}$) which resembles the signature of 
an edge-on rotating disc \citep*{Kregel04}, which is what we take it to be. 
Gas further to the North and in particular to the South could either be part of
the interaction or be clumps further out in the disc. Note the lack of 
connection between galaxy `C' and VIRGOHI 21.

\begin{figure*}
\begin{minipage}{126mm}
\includegraphics[width=126mm]{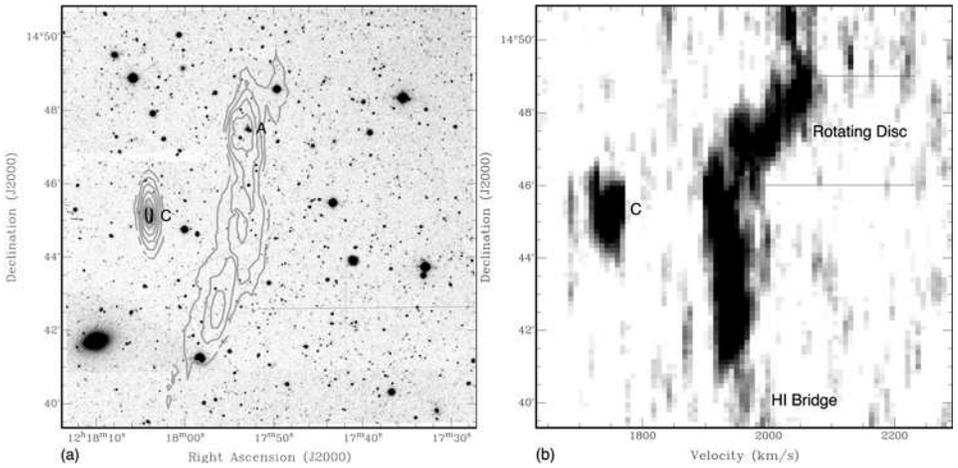}
\caption{As Fig. 1, but expanded and superimposed on a negative of our deep
CCD $B$-band image \citep{Minchin05} with a surface-brightness limit of 27.5 
$B$ mag. arcsec$^{-2}$.}
\end{minipage}
\end{figure*}

\section{Discussion}

Combining the new data with the old, we argue:

(a) If attributed to gravitation, changes in velocity of galactic size over
galactic scales, as seen here, require masses of galactic proportions.  On
dimensional grounds the velocity wrench $\Delta V$ ($\sim 200$ km\,s$^{-1}$)
seen at VIRGOHI 21 over a conservative length-scale $\Delta x$ ($\sim 14$ kpc 
$\simeq 5\times 10^{22}$ cm $\simeq 3$ arcmin at the Virgo Cluster distance
of 16 Mpc \citep{Minchin05})
implies a mass $M \geq (\Delta V)^2 \Delta x/G \simeq 10^{10-11}$ solar masses.
More specifically the gravitational (free-fall) timescale 
$(\Delta x)^{3/2}/(GM)^{1/2}$
needs to be $\leq\Delta x/V$, the time it takes for gas travelling at $V$
(relative to $M$) to change velocity by $\Delta V$.  And where $\Delta V \sim
V$, as it appears to be here (see Fig. 2b), $M\geq (\Delta V)^2 \Delta x/G$
again.  For non-gravitational alternatives see below.

(b) The bridge between NGC 4254 and VIRGOHI 21 reveals that they have 
interacted.  According to published numerical simulations \citep{Vollmer05} 
(below) the morphological peculiarities of NGC 4254 require a perturbing mass 
of $\sim 10^{11} M_\odot$ -- providing independent evidence that
VIRGOHI 21 weighs $\sim 10^{11} M_\odot$.  The simulation implies an 
interaction $3\times 10^8$ years ago.  As the projected length of the bridge
is 120 kpc this would imply it has been drawn out at a projected speed of 390 
km\,s$^{-1}$, which is comparable to the radial velocity difference 
between NGC 4254 and VIRGOHI 21 of 400 km\,s$^{-1}$.

(c) NGC 4262, the spiral to the North-East, is not involved.  There is no 
bridge to it (see animation) and the
radial velocity difference between NGC 4254 and NGC 4262 is too large
(900 km\,s$^{-1}$) to generate one.

(d) If object C, the H\,{\sc i} galaxy just to the east (left) of VIRGOHI 21 
(Fig. 2) were involved its mass, by (a), must be $\sim 10^{11} M_\odot$.  
However 
its measured H\,{\sc i} velocity distribution and size suggest a probable mass 
$\leq 10^9 M_\odot$, while it has a luminosity of only $10^8 L_\odot$ and a $
M_{HI}$ of $4\times 10^6 M_\odot$.  It is two orders of magnitude too 
undermassive and underluminous to explain the dynamics of VIRGOHI 21, 
while there is no sign of its interacting.

(e) The structure of VIRGOHI 21 (Fig. 2a) is centred at R.A. $12^h17^m52^s$ and
elongated in the North-South plane between $+14^\circ41^\prime$ and 
$+14^\circ49^\prime$.  Its remarkable velocity change (from 1900 
to 2100 km\,s$^{-1}$) is, however, confined to the tilted structure (Fig. 2b)
between $+14^\circ46^\prime$ and $+14^\circ49^\prime$, which closely resembles
the characteristic shape of an edge-on disc \citep{Kregel04}.  If indeed what
we are witnessing here is a dark, gravitationally bound, edge-on rotating disc
then its properties are as presented in Table \ref{properties}, with a minimum
mass of $2\times 10^{10} M_\odot$.  Judging from visible disc galaxies, whose
masses continue to rise beyond their H\,{\sc i} edges \citep{Salucci97}, the 
full size and mass of such a disc could easily reach $\sim 10^{11} M_\odot$.
The very low surface-brightness limits 
(dimmer than 27.5 $B$ mag arcsec$^{-2}$ \citep{Minchin05}) imply that the
disc has a $M_{dyn}/L_B$ ratio of at least 750 $M_\odot/L_\odot$
where normal galaxies have $< 50$.  
Deep HST observations, capable of reaching individual Red Giant 
stars in VIRGOHI 21, are in hand.

\begin{table}
\caption{PROPERTIES OF THE DARK DISC}
\label{properties}
\begin{tabular}{ll}
Diameter $2R$, [from $+14^\circ 46^\prime$ to $49^\prime$] & 14 kpc\\
Circular Velocity $V_c$, [$(2100 - 1900)/2$] & 100 km\,s$^{-1}$\\
Spin Period $P$, [$2\pi R/V_c$] & $4\times 10^8$ years\\
Total Mass $M_T$, [$RV_c^2/G$] & $2\times 10^{10} M_\odot$\\
Face-on Mass-density [$M_T/\pi R^2$] & $2\times 10^{-2}$ g cm$^{-2}$\\
Hydrogen Mass $M_{HI}$, [$F_{HI} = 0.7$ Jy\,km\,s$^{-1}$]& $4 \times 10^7
M_\odot$\\
Face-on gas density $N_{HI}$, [$M_{HI}/\pi m_{H}R^2$] & $3\times 10^{19}$
cm$^{-2}$\\
Total Mass to Blue Light Ratio $M_{HI}/L_B$& $> 750 M_\odot/L_\odot$\\
\multicolumn{2}{l}{[assuming a disc $0.5^\prime \times 3^\prime$]}
\end{tabular}
\end{table}

The integrated spectrum, minus the bridge, is consistent with earlier 
single-dish observations, implying that the interferometer misses little 
H\,{\sc i}.  The surprising ease with which it has 
been mapped is due to two pieces of luck: it is edge on, thus increasing the
apparent surface density, and most of its gas is spread over fairly low
velocity widths within each synthesised beam.

The incontrovertible evidence (see animation) of an interaction with NGC 4254
provides independent support for the massive nature of VIRGOHI 21.
NGC 4254 is a luminous one-armed spiral galaxy sufficiently peculiar to have
attracted several studies (\citealt*{Iye82,Phookun93};\citealt{Vollmer05}).  
Single-armed  spirals are invariably the result of interactions with close-by 
massive companions \citep{Iye82}.
The lack of any visible companion thus triggered
observations and dynamical models.  Recent numerical models by
\citet{Vollmer05} indicate that NGC 4254 ``had a close and rapid 
encounter
with a $10^{11} M_\odot$ galaxy $\sim 250$ Myr ago.  The tidal interaction 
caused the spiral structure...''.  \citet{Phookun93}, in a VLA 
study of
the galaxy, find a trail of gas leading away from it (their Figure 5) in both
the right direction and with exactly the right velocity-gradient required
to intersect VIRGOHI 21.  Thus the case for VIRGOHI 21 being the aforesaid
$\sim 10^{11} M_\odot$ mass which caused the peculiarities in NGC 4254 seems 
strong.  A detailed simulation which includes our new data could remove 
any doubt.

Models for VIRGOHI 21 now have a number of crucial observations to explain: 
the broad velocity-width in a galactic volume (implying large mass);
its elongated geometry; its steep velocity profile; the bridge to NGC 4254, 
and to no other galaxy; the damage to NGC 4254 (large mass); and the lack of 
light.  Given the above observations we now discuss a number of hypotheses
as to the origin of VIRGOHI 21 and its associated H\,{\sc i}.

{\bf (a) We are detecting tidal debris left by the past interaction of NGC 4254
and another galaxy.}

This is rather
easy to dismiss here where the lines are so broad, and there are no such
interactors visible in the immediate vicinity.
Imagine two galaxies with radial velocities, $V_1$ and $V_2$ at either end
of an approximately linear tidal bridge of physical length $d$ pitched at an 
angle $\theta$ to the plane of the sky.  A telescope pointed towards it has a 
transverse beam diameter of $b$ at the bridge.  The only significant
gas motions within the bridge will be streaming velocities along its length.
From end to end of the bridge the radial velocity difference is $|V_2 - V_1|$
while within the telescope beam the measured velocity width, $\Delta V$, will 
be ($b/d\sin\theta)\times|V_2 - V_1|$.  But, as is well known, bridges of any 
size arise only when the {\it total} velocity difference between the 
interacting galaxies -- $|V_2 - V_1|/\cos\theta$ here -- is of the same order 
as the circular velocity $V_c$ in the donor \citep{Toomre72}.  It 
follows immediately that $\Delta V/V_c \simeq (b/d)\tan\theta$.  Thus broad
line widths $\Delta V \simeq V_c$, as here, can only be seen within a beam if
{\it both} interactors {\it appear} to lie within, or very close to the beam
($d\cos\theta\leq b$).  There are no such putative interactors even within the 
Arecibo beam ($b\simeq 3\farcm6$) or close to it, 
otherwise we would see them in Fig. 2a.  VIRGOHI 21 cannot be 
such tidal debris. And one cannot escape this conclusion by presupposing a
very ancient interaction almost along the line-of-sight ($\theta \sim 
90^\circ$) -- if so, where is the culprit?

\citet*{Bekki05b} have carried out numerical simulations to model interactions
that might lead to tidal features detectable at 21 cm.  They
have proposed an interaction with NGC 4254 as the likely cause of VIRGOHI 21.
Their Model 1, which does create a cloud of the right velocity width -- because
of its projection along the line of sight, fails to show the perturbing galaxy,
and in any case would 
never have been claimed as a plausible dark-galaxy candidate by us because it 
would fail the stringent `timing-argument' explained in our previous 
paper \citep{Minchin05}.  Additionally, the sense of their velocity field 
relative to the donor, NGC 4254, is opposite to the sense actually observed.  
Their favoured Model 4, which does indeed recreate the disturbance of NGC 
4254, has a velocity width of 20, not 200, km\,s$^{-1}$!  In fact, their 
simulations demonstrate just how hard it is to explain VIRGOHI 21 as tidal
debris.

Other hypotheses have also been put forward, and should be examined to see if
they can explain our observations:

{\bf (b) Two superposed H{\sc i} Clouds.}

The components of VIRGOHI 21 are connected both spatially and in
velocity, making it exceedingly improbably that they could be chance
superpositions of clouds, while the bridge to NGC 4254 is not explained by
this hypothesis.

{\bf (c) A tidal tail from a galaxy merging with NGC 4254.}

\citet{Phookun93} suggested that the distortion of NGC 4254 could be due
to infalling gas-clouds.  Could VIRGOHI 21 be a tail, similar to those seen
in UGC 10214 \citep{Briggs01} or NGC 4038/9 \citep*{Gordon01}?  In
NGC 4038/9, in particular, there is what appears to be a tidal dwarf  near 
a strong concentration of H\,{\sc i} at the end of the southern tail, around 
60 kpc from the centre of the system (for a distance of 13.8 
Mpc; \citealt*{Saviane04}).

Although this hypothesis can, at first sight, explain the WSRT observations,
with the bridge as an H\,{\sc i} tail and VIRGOHI 21 as a tidal dwarf forming
in a concentration at the tip, it is clear on more detailed examination that 
this
cannot be the case.  The velocity field in VIRGOHI 21 changes direction in a 
way not seen in examples of tidal dwarfs, and to a much greater extent -- the 
tidal dwarf
in the NGC 4038/9 system, for instance, has a gradient in the same sense as the
gas in the tail nearby, whilst VIRGOHI 21's gradient is in the opposite 
direction and has twice the velocity width.  Even more troubling for this 
hypothesis, NGC 4254 does not show signs of having recently merged with another
$L^\star$ sized galaxy -- and yet it would have had to do so to have thrown out
a tidal tail
twice as long as that from the violently interacting NGC 4038/9 system.

{\bf (d) A high column-density part of a giant H\,{\sc i} ring.}

\citet{Bekki05a} propose that some H\,{\sc i} clouds without optical 
counterparts
could be the high-density regions of H\,{\sc i} rings.  In their scenario,
objects such as the Leo Ring \citep{Schneider83} are formed by the tidal
stripping of low surface-brightness galaxies with extended gas disks.  
As the column-density varies around the ring, it is possible that only part
of it will be detectable -- possibly at a large distance from the original
galaxy from which the gas has been stripped.  This would then be identified
as an intergalactic H\,{\sc i} cloud.

While this hypothesis might have explained the single-dish observations, it is 
hard to
make the WSRT data fit.  The ring should be orbiting around a large centre of 
mass to the east of VIRGOHI 21 and at a higher velocity, which cannot be 
identified.  Also, if the bridge is part of the ring, then it is unconnected 
with the perturbation of NGC 4254 -- which is both very unlikely and leaves
the single-arm mode of that galaxy unexplained.

{\bf (e) A three-body interaction.}

It might be that a third mass, possibly object C, has 
interfered with the interaction between NGC 4254 and another body (possibly
NGC 4262, as proposed by \citet{Vollmer05}, as there are no other 
obvious candidates).  However, it seems very unlikely that C has sufficient
mass to cause a perturbation of over 200 km\,s$^{-1}$ to the tidal stream,
and VIRGOHI 21 passes to its west, not between it and NGC 4262 as might
be expected if it had pulled the stream westward.  The radial velocity
of C means that it must be moving past the stream at a velocity
(relative to the putative undisturbed velocity of the stream at that point
of 2100 km\,s$^{-1}$ -- the velocity of the high-velocity end of VIRGOHI 21)
of at least 350 km\,s$^{-1}$.  At this speed, it would not have stayed close
to the stream long enough to have severely disturbed it.

It would also be expected
that if C were involved in the interaction then there would be gas
falling onto it, but this does not appear to be the case as there is no gas
seen between VIRGOHI 21 and C.  Nor does C show any signs of 
being disturbed itself; its H\,{\sc i} is unresolved, implying that it is 
confined to the area of the optical galaxy, and its optical image similarly
shows no sign of any perturbation.

{\bf (f) Ram pressure stripping.}

\citet{Oosterloo05} argue that another H\,{\sc i} cloud in the 
Virgo Cluster, VIRGOHI 4 \citep{Davies04},
is caused by ram-pressure stripping from NGC 4388 due to an interaction with
the hot-gas halo of the M86 sub-group and suggest a similar origin could
be possible for VIRGOHI 21.  This would explain the bridge without the need
to invoke a second galaxy, either interacting or merging.  However, this
cannot give the steep, reversed velocity gradient seen in VIRGOHI 21, nor does 
it explain the distortion to the optical disk of NGC 4254.  Ram pressure 
stripping does not, therefore, appear to work as an explanation of this system.

{\bf (g) Harassment.}

Harassment occurs in much faster interactions than tidal 
stripping \citep*{Moore98}, thus it
offers a possible mechanism for removing gas from NGC 4254 while leaving the
interactor sufficiently distant that it can not be easily identified.  However,
such an interaction, while it might remove gas, would not form the tidal bridge
seen here between VIRGOHI 21 and NGC 4254.

Our conclusion is that the most plausible explanation is that VIRGOHI 21 is
the cause of the tidal disturbance not the effect of other physical processes.

As might be expected for a dark galaxies, VIRGOHI 21 has a shortage of baryons
compared to visible galaxies.  Visible disc galaxies typically have,
in the form of stars and gas, $\sim 10$ per cent of their dynamical mass
in baryons, a figure consistent with cosmological models which predict
$\sim 7$ times as much Dark Matter as baryonic in the Universe at large
\citep {Salucci97}.  Had VIRGOHI 21 10 per cent of its mass in baryons it would
contain
$\sim 2\times 10^9$ solar masses which would, if it were once in the form
of H\,{\sc i} spread evenly across the existing dynamical disc, yield a
column density $\sim 10^{21}$ atoms cm$^{-2}$ -- sufficiently above
the ``star-formation threshold'' of $\sim 10^{20}$ cm$^{-2}$ \citep{Martin01},
or above the Toomre criterion for gravitational instability in a rotating 
disc \citep{Toomre64}, to form stars \citep{Jimenez97,Verde02}.  Since it 
apparently contains no stars it must have {\it and must always have had}, ten 
times less baryons -- a phenomenon often 
apparent in dwarf galaxies \citep{Mateo98}.  Thus
90 per cent of the baryons one might have expected to find in a typical galaxy
appear to be missing.  Whatever the reason, the lack of baryons in such
massive objects can only decrease their detectability.

\section{Conclusions}

Models for VIRGOHI 21 now have a number of crucial observations to explain: the
broad velocity-width in a galactic volume (implying large mass); its elongated 
geometry; its steep velocity profiles; the bridge to NGC 4254, and to no other
galaxy; the damage to NGC 4254 (large mass); and the lack of light. Of other 
models that have been suggested, or that we have been able to devise, neither 
tidal debris, nor Ram Pressure Stripping nor a Group gravitational field, nor 
long-range interactions deliberately aimed at simulating this  object can 
explain the abrupt changes of velocity. 

The new observations make it even harder to 
escape the inference that VIRGOHI 21 contains a massive dark disc.

\section*{acknowledgements}
We thank the Netherlands Foundation for Radio Astronomy for use of the
WSRT, the National Astronomy and Ionosphere Center for use of the Arecibo
Telescope, the UK Particle Physics and Astronomy Research Council for financial
support, and the Australia Telescope National Facility for inspiration and
technical support.  We would like to thank the following for useful comments:
Virginia Kilborn, Erwin de Blok, Martin Zwaan and Greg Bothun.

\end{document}